\documentclass[12pt]{article}

\usepackage{epsfig}
\usepackage{amstex}
\usepackage{amssymb}
\usepackage{axodraw}

\begin{document}

\marginparwidth 3cm
\setlength{\hoffset}{-1cm}
\newcommand{\mpar}[1]{{\marginpar{\hbadness10000%
                      \sloppy\hfuzz10pt\boldmath\bf\footnotesize#1}}%
                  \typeout{marginpar: #1}\ignorespaces}
\def\mda{\mpar{\hfil$\downarrow$\hfil}\ignorespaces}
\def\mua{\mpar{\hfil$\uparrow$\hfil}\ignorespaces}
\def\mla{\marginpar[\boldmath\hfil$\rightarrow$\hfil]%
                   {\boldmath\hfil$\leftarrow $\hfil}%
                   \typeout{marginpar:
                     $\leftrightarrow$}\ignorespaces}

\renewcommand{\abstractname}{Abstract}
\renewcommand{\figurename}{Figure}
\renewcommand{\thefootnote}{\fnsymbol{footnote}}

\providecommand{\SP}{\scriptstyle}
\providecommand{\ns}{\hspace{2mm}}

\providecommand{\hp}{H^+}
\providecommand{\hm}{H^-}
\providecommand{\tanb}{\text{tan}\beta}
\providecommand{\cotb}{\text{cot}\beta}
\providecommand{\sina}{\text{sin}\alpha}
\providecommand{\cosa}{\text{cos}\alpha}
\providecommand{\sinb}{\text{sin}\beta}
\providecommand{\cosb}{\text{cos}\beta}
\providecommand{\shat}{\hat{s}}
\providecommand{\that}{\hat{t}}
\providecommand{\uhat}{\hat{u}}
\providecommand{\bhat}{\hat{\beta}}
\providecommand{\thone}{\hat{t}_1}
\providecommand{\uhone}{\hat{u}_1}
\providecommand{\epone}{\epsilon_1}
\providecommand{\eptwo}{\epsilon_2}
\providecommand{\qslash}{\not \!q}
\providecommand{\koneslash}{{\not \!k}_1}
\providecommand{\ktwoslash}{{\not \!k}_2}
\providecommand{\pthrslash}{{\not \!p}_3}
\providecommand{\qslashsmall}{\not \,q}
\providecommand{\kslash}{{\not \!k}}
\providecommand{\pslash}{{\not \!p}}
\providecommand{\koneslashsmall}{{\not \,k}_1}
\providecommand{\ktwoslashsmall}{{\not \,k}_2}
\providecommand{\pthrslashsmall}{{\not \,p}_3}
\providecommand{\intn}{\int \!d^n q}
\providecommand{\gf}{G_{\text{F}}}
\providecommand{\as}{\alpha_{\text{s}}}

\begin{titlepage}

\begin{flushright}
CERN-TH/97-137 \\
DESY 96-254 \\
hep-ph/9707430 \\
July 1997
\end{flushright}

\vspace{1cm}

\begin{center}
\baselineskip25pt

\def\thefootnote{\fnsymbol{footnote}}
{\large\sc Production of Charged Higgs Boson Pairs \\ 
           in Gluon--Gluon Collisions}

\end{center}

\setcounter{footnote}{3}

\vspace{1cm}

\begin{center}
\baselineskip12pt

{
A.~Krause$^1$,
T.~Plehn$^1$,
M.~Spira$^2$
and P.~M.~Zerwas$^1$}  

\vspace{1cm}

$^1$ {\it Deutsches Elektronen--Synchrotron DESY, D--22603 Hamburg, FRG}
\\[5mm]
$^2$ {\it Theory Division, CERN, CH--1211 Geneva 23, Switzerland}

\vspace{0.3cm}

\end{center}

\vspace*{\fill}

\begin{abstract}
  The search for charged
  Higgs bosons, which are predicted in supersymmetric theories, is
  difficult at hadron colliders if the mass is large. In this paper
  we present the theoretical set-up for the production of charged
  Higgs boson pairs at the LHC in gluon--gluon collisions:
  $pp\!\rightarrow \! gg\!\rightarrow \!\hp \hm$. When established
  experimentally, the trilinear
  couplings between charged and neutral CP-even Higgs bosons,
  $\hp\hm h^0$ and $\hp\hm H^0$, can be measured.
\end{abstract}

\vspace*{\fill}

\begin{flushleft}
CERN-TH/97-137 \\
DESY 96-254 \\
hep-ph/9707430 \\
July 1997
\end{flushleft}

\end{titlepage}

\def\thefootnote{\arabic{footnote}} \setcounter{footnote}{0}

\setcounter{page}{2}

\section{Introduction}
Charged Higgs bosons are part of the extended Higgs sector predicted
in supersymmetric theories.  They belong to the two Higgs isodoublets
that must be introduced to provide masses to down- and up-type
fermions through the superpotential and to keep the theory free of
anomalies \cite{Fayet}.  The mass of these particles is expected to be
in the range of the electroweak symmetry-breaking scale $v=246$\,GeV,
albeit with a rather wide spread. Embedding low-energy supersymmetry
into supergravity models with universal SUSY breaking, the masses of
the heavy neutral and charged Higgs bosons are indeed predicted at the
TeV scale for large parts of the parameter space [see
e.g.~Ref.\,\cite{Barger}].

  At hadron colliders \cite{Richter}, several mechanisms can be
  exploited to search for charged Higgs bosons. A copious source
  of charged Higgs bosons with fairly light mass are decays of
  top quarks, $t\rightarrow b+\hp$. Since top quarks are produced with
  very large rates at the LHC, charged Higgs bosons can be searched
  for in this channel for masses $m_{H^\pm}$ up to the top mass.
  No experimental technique has
  been established so far for charged Higgs
  bosons with masses above the top mass.\footnote{At $e^+e^-$
  colliders, charged Higgs bosons can be searched for in the process
  $e^+e^-\rightarrow \hp\hm$ for masses up to the beam
  energy, {\it i.e.~}up
  to $m_{H^\pm} \lesssim 1$\,TeV for prospective 2\,TeV $e^+e^-$ linear
  colliders.}
  Channels such as $pp \rightarrow \bar tb\hp$, in which the Higgs bosons
  are emitted from heavy quark lines \cite{Moretti}, and the Drell--Yan
  production of charged
  pairs \cite{Eichten}, $pp\rightarrow q\bar{q}\rightarrow \hp\hm$, are
  potential
  candidates for the search.

  Extending the previous analysis of neutral
  Higgs bosons in
  supersymmetric theories, Ref.\,\cite{Plehn}, we present in this
  paper the theoretical set-up for the production of charged Higgs boson
  pairs
  in gluon--gluon collisions:
  \begin{equation}
      pp\rightarrow gg\rightarrow \hp\hm \; .
  \end{equation}
  The analysis of this channel is motivated by the large number of
  gluons in high-energy proton beams.  We work out the predictions for
  the cross sections\footnote{The triangle contributions to the
    production cross section have already been determined in
    Ref.\,\cite{Willenbrock}, but the more complicated box
    contributions were only estimated in this
    paper.}$^,$\footnote{After finalizing this calculation we received
    a copy of the paper \cite{Yi}.  The present analysis differs from
    this paper in several aspects: (i) We present the partonic cross
    sections in compact analytical form, while they are given in
    Ref.\,\cite{Yi} in terms of twelve form factors from which the
    individual UV singularities are not separated yet.  (ii) We
    include a careful analysis of the theoretically interesting large
    quark-mass limit, which serves as an important cross-check of the
    calculation.  (iii) Our final results differ from those in
    Ref.\,\cite{Yi}: Part of the discrepancies could be traced back to
    a wrong flux factor and a wrong factor of 2 in the gluon
    luminosity of \cite{Yi} (moreover, opposite to the calculation in
    Ref.\,\cite{Yi}, $Z$-exchange cannot contribute to the $gg$ fusion
    to charged Higgs boson pairs as a result of symmetry arguments);
    other points could not be isolated in detail since the cross
    sections in \cite{Yi} are evaluated numerically in a global way
    that does not allow checks of separate terms. The results
    presented in our paper have been cross-checked in two independent
    calculations.}  at the LHC. For charged Higgs masses above the top
  mass, yet less than 300\,GeV, the cross sections will turn out to be
  in the range between 1\,fb and 10\,fb if the mixing parameter
  $\tanb$ in the 2-doublet Higgs sector is either small, $\tanb
  \gtrsim 1$, or large, $\tanb \sim 50$. The search for such events
  will therefore be difficult, and the characteristic decay properties
  must be exploited exhaustively in the experimental
  analyses\footnote{Such an experimental analysis is beyond the scope
    of the present paper, which is restricted to the theoretical basis
    of the gluon-fusion process.}. However, if the signal can be
  established experimentally, the trilinear couplings between charged
  and neutral CP-even Higgs bosons, $h^0\,\hp\,\hm$ and
  $H^0\,\hp\,\hm$, can be studied, thus complementing related analyses
  in the purely neutral sector \cite{Plehn,Djouadi}. Measurements of
  these couplings are the first step in the reconstruction of the
  self-interaction terms in the Higgs potential.  By contrast, the
  emission of charged Higgs bosons from top quarks and the Drell--Yan
  production of charged Higgs pairs do not include the trilinear
  couplings.

  The paper is organized as follows. After a brief summary of the
  charged Higgs sector within
  the Minimal Supersymmetric Standard Model (MSSM) in the next
  section, the theoretical set-up for the production of charged Higgs
  boson pairs in gluon--gluon collisions at the LHC will be described
  in the subsequent section which includes a brief phenomenological
  evaluation. 
  
\section{Charged Higgs Bosons: SUSY Expos\'{e}}
  The MSSM will be adopted as the paradigm for a theory of
  charged Higgs bosons. After the
  characteristic phenomena are worked out for this example, the results
  can easily be adapted to more general scenarios. At least two Higgs
  isodoublets are required in supersymmetric theories to generate masses
  for down- and up-type
  fermions through the trilinear superpotential, and to cancel
  anomalies associated with higgsino triangle loops. After
  absorbing three of the eight field components to generate the
  longitudinal components of the electroweak gauge bosons, three
  neutral CP-even and CP-odd particles and a pair
  of charged particles build up the Higgs spectrum of the
  MSSM \cite{Fayet}.
  
  In the minimal version, the properties of the Higgs particles depend
  essentially on three parameters: two masses, generally identified
  with the $Z$-boson mass $m_Z$ and the pseudoscalar Higgs-boson mass
  $m_A$, and the ratio of the two vacuum expectation values of the
  neutral Higgs fields, $\tanb=v_2/v_1$, assumed to vary between unity
  and $m_t/m_b\sim 50$. Including the leading radiative corrections,
  induced in the potential by the large top mass, the other masses can
  be expressed in terms of $m_Z$ and $m_A$:
  \begin{alignat}{3}
       m^2_{h^0,H^0} &= \frac{1}{2}\Big[ m_{AZ}^2
                       \mp \sqrt{ m_{AZ}^4-4m_A^2 m_Z^2\,
                     {\text{cos}}^2\left( 2\beta \right)
                     -4\epsilon \left( m_A^2 \,\text{sin}^2\beta
                                     +m_Z^2 \,\text{cos}^2\beta \right) }
                                \Big] 
       \qquad\phantom{ss} \\
       m^2_{H^\pm} &= m_A^2+m_W^2 
  \end{alignat}
  with the abbreviation $m_{AZ}^2 = m_A^2+m_Z^2+\epsilon\,$; the
  leading corrections are characterized by the radiative parameter
  $\epsilon$ \cite{Hempfling},
  \begin{equation}
      \epsilon = \frac{3 G_F}{\sqrt{2} \pi^2 }
                 \frac{m_t^4}{{\text{sin}}^2 \beta}
                 \text{log} \left[\frac{m_{\tilde{t}}^2}{m_t^2} \right]
      \; .
  \end{equation}
  The parameter $m_{\tilde{t}}^2=m_{\tilde{t}_1}m_{\tilde{t}_2}$ denotes
  the average mass square of the stop
  particles. The
  mixing in the
  CP-even Higgs sector is given by the angle $\alpha$ with  
  \begin{equation}
      \text{tan} \left( 2 \alpha\right) = 
                 \frac{m_A^2 + m_Z^2}{m_A^2 - m_Z^2 + 
                       \epsilon/ \text{cos} \left( 2\beta \right)}
                 \, \text{tan} \left( 2 \beta\right)
      \quad\qquad \left[\,-\frac{\pi}{2} \le \alpha \le 0\,\right]
      \; .
  \end{equation}
  The mass pattern is a consequence of the fact that the size of the
  quartic self-couplings is determined by the square of the gauge
  couplings in the MSSM. Since these are small, strong
  upper bounds on the mass of the lightest neutral CP-even Higgs boson
  $h^0$ can be derived.
  The lightest neutral Higgs boson $h^0$ may therefore be the first
  Higgs particle to be discovered. However, since
  it is very difficult to determine $\tanb$, the measurement of the $h^0$
  mass will
  in general not enable us to predict the mass of the charged Higgs
  bosons. Moreover,
  as a result of the decoupling theorem, $m_{h^0}$ becomes
  increasingly insensitive to the values of the heavy Higgs masses
  beyond $\sim$ 200\,GeV, as is evident from Fig.~1.
  \begin{figure}[h]
  \begin{center}
  \epsfig{file=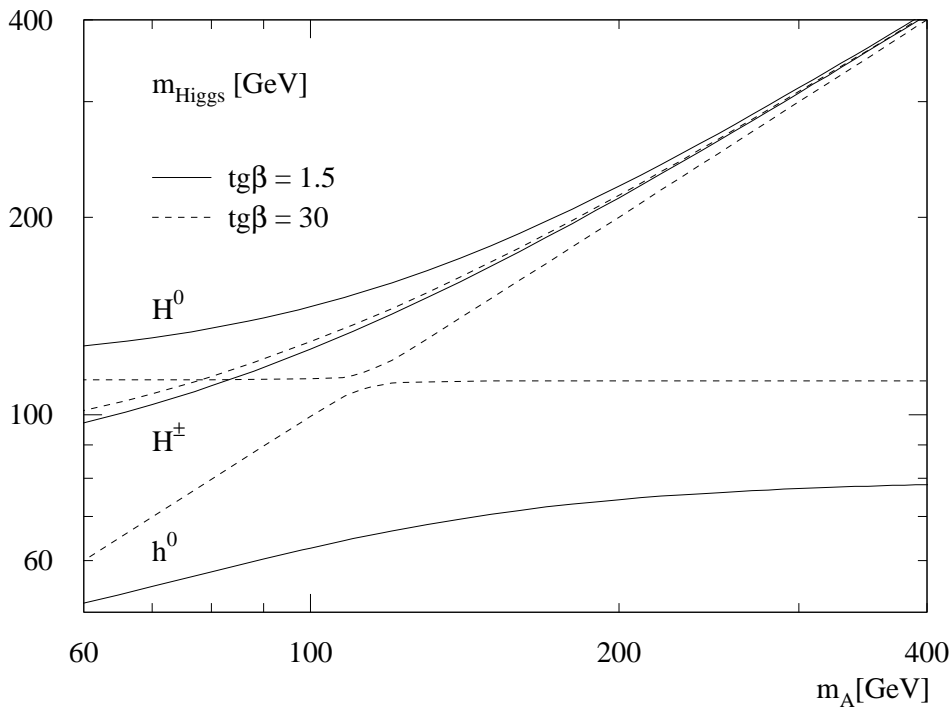,width=13cm,angle=0}
  \caption{\it The masses of the Higgs bosons in the MSSM for 
    two representative values of $\tanb=1.5$ and $30$.}
  \end{center}
  \end{figure}

  \begin{small}
  \begin{table}
  \renewcommand{\arraystretch}{1.5}
  \begin{center}
  \begin{tabular}{|l||l|c|}\hline \rule[0mm]{0mm}{6.5mm}
        SM & $g_t$ & ${\left( \sqrt{2} G_F \right)}^{\frac{1}{2}} 
             m_t$ \\\rule[-4mm]{0mm}{11.5mm}
           & $g_b$ & ${\left( \sqrt{2} G_F \right)}^{\frac{1}{2}} 
             m_b$ \\ \hline\hline
        \rule[2mm]{0mm}{5.5mm}
        \!\!MSSM & $\bar tth^0$ & $\left( \text{cos}\alpha/\text{sin}\beta 
                                            \right) g_t$  \\ 
        & $\bar bbh^0$ & $-\left( \text{sin}\alpha/\text{cos}\beta 
                                 \right) g_b$ \phantom{ss} \\
        & $\bar ttH^0$ & $\left( \text{sin}\alpha/\text{sin}\beta 
                                            \right)  g_t$  \\
        & $\bar bbH^0$ & $\left( \text{cos}\alpha/\text{cos}\beta
                                            \right)  g_b$  \\
        & $\bar tbH^+$ & $-\frac{1}{\sqrt{2}} 
             \left[ g_t\, \text{cot}\beta\left( 1-\gamma_5 \right)
                   +g_b\, \text{tan}\beta\left( 1+\gamma_5 \right)
             \right]$ \\
        \rule[-4mm]{0mm}{8mm} 
        & $\bar btH^-$ & $-\frac{1}{\sqrt{2}} 
             \left[ g_t\, \text{cot}\beta\left( 1+\gamma_5 \right)
                   +g_b\, \text{tan}\beta\left( 1-\gamma_5 \right)
             \right]$ \\
        \hline
  \end{tabular}
  \end{center}
  Table 1: {\it The Higgs-quark couplings to fermions in the MSSM
                    relative to the corresponding SM couplings.}
  \renewcommand{\arraystretch}{1.2}
  \end{table}
  \end{small}
  
  The couplings of the SUSY Higgs particles
  to fermions are
  modified
  by the mixing angles with respect to the
  Standard Model (Table 1). They are renormalized only indirectly
  in leading order
  through the renormalization of the mixing angle 
  $\alpha$.
  By contrast, the trilinear Higgs couplings are
  modified indirectly
  through the renormalization of the mixing angle $\alpha$, but
  also explicitly through vertex corrections \cite{Haber}.
  Normalized in units of $\lambda_0 = 
  {\left( \sqrt{2} G_F \right)}^{\frac{1}{2}} m_Z^2$, they are given by
  \begin{alignat}{3}
       \lambda (h^0H^+H^-) &= 2 \, {\text{cos}}^2 \theta_W \, 
                           \text{sin}\left(\beta -\alpha\right) +
                           \text{cos}\left( 2\beta\right)
                           \text{sin}\left(\beta +\alpha\right)
                          +\frac{\epsilon}{M_Z^2}
                           \frac{\text{cos}\alpha\,
                                 {\text{cos}}^2\beta}
                                {\text{sin}\beta} 
       \label{eq:lambda1} \\
       \lambda (H^0H^+H^-) &= 2 \, {\text{cos}}^2 \theta_W \,
                           \text{cos}\left(\beta -\alpha\right) -
                           \text{cos}\left( 2\beta\right)
                           \text{cos}\left(\beta +\alpha\right) 
                          +\frac{\epsilon}{M_Z^2}
                           \frac{\text{sin}\alpha\,
                                 {\text{cos}}^2\beta}
                                {\text{sin}\beta}       
       \label{eq:lambda2} \; .
  \end{alignat}
  Their size
  is illustrated for $\tanb=1.5$ and 30 in Fig.~2. 
  \begin{figure}[h]
  \begin{center}
  \epsfig{file=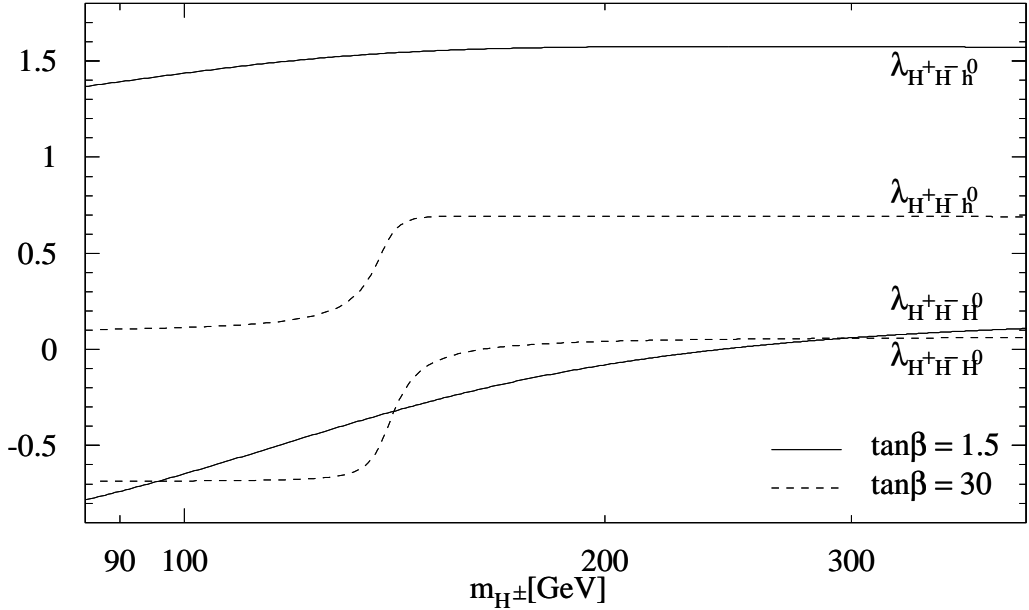,width=14cm,angle=0}
  \caption{\it The trilinear couplings of the charged Higgs
    bosons in the MSSM for two representative values of $\tanb=1.5$
    and $30$.}
  \end{center}
  \end{figure}
  The trilinear couplings of the light scalar Higgs boson to charged
  Higgs particles turn out to be positive in the relevant
  charged-Higgs mass range, while those of the heavy scalar Higgs boson are
  negative for small charged Higgs masses and change to positive but
  small values for
  charged Higgs masses between about 150 and 250 GeV. The absolute size of
  these couplings extends from ${\cal O}(10^{-1})$ to ${\cal O}(1)$.
  For large values of $\tanb$ the trilinear couplings change rapidly
  near $m_{H^\pm}\sim 150$~GeV. 
    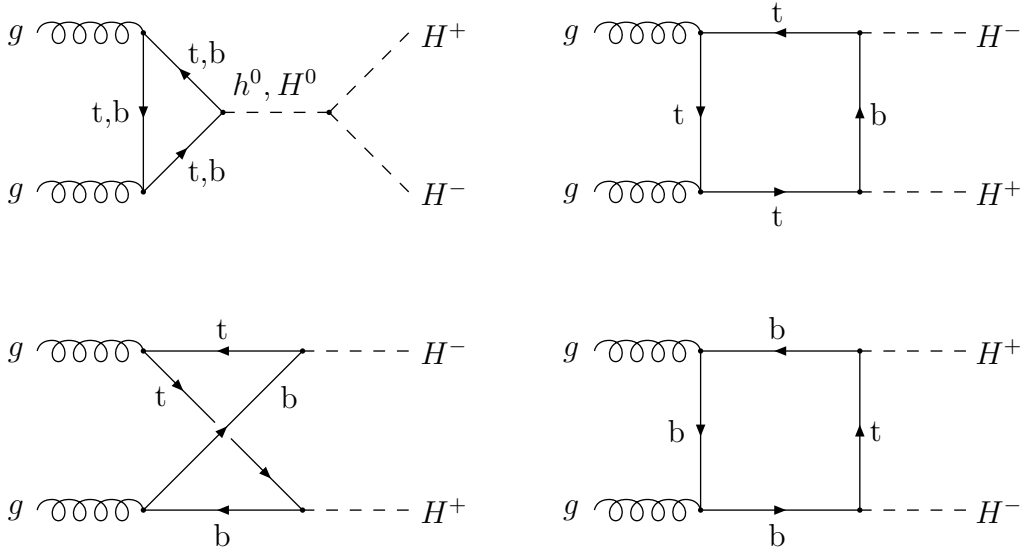
\begin{figure}[h]
    \begin{center}
    \begin{picture}(350,220)(0,0)
        \Gluon(0,170)(40,170){4}{4} \Text(-5,170)[r]{$g$} 
        \Vertex(40,170){1}
        \Gluon(0,230)(40,230){4}{4} \Text(-5,230)[r]{$g$}
        \Vertex(40,230){1}
        \ArrowLine(40,230)(40,170) \Text(35,200)[r]{t,b}
        \ArrowLine(70,200)(40,230) \Text(57,217)[lb]{t,b}
        \ArrowLine(40,170)(70,200) \Text(57,183)[lt]{t,b}
        \Vertex(70,200){1}
        \DashLine(70,200)(110,200){5} \Text(90,205)[b]{$h^0,H^0$}
        \Vertex(110,200){1}
        \DashLine(110,200)(140,230){5} \Text(145,230)[l]{$H^+$}
        \DashLine(140,170)(110,200){5} \Text(145,170)[l]{$H^-$}

        \Gluon(0,50)(40,50){4}{4} 
                      \Text(-5,50)[r]{$g$} 
        \Vertex(40,50){1}
        \Gluon(0,110)(40,110){4}{4} \Text(-5,110)[r]{$g$}
        \Vertex(40,110){1}
        \ArrowLine(40,110)(67,83) \Text(49,97)[rt]{t}
        \ArrowLine(73,77)(100,50) 
        \ArrowLine(100,50)(40,50) \Text(70,45)[t]{b}
        \ArrowLine(40,50)(100,110) \Text(92,97)[lt]{b}
        \ArrowLine(100,110)(40,110) \Text(70,115)[b]{t}
        \Vertex(100,50){1}
        \Vertex(100,110){1}
        \DashLine(100,110)(140,110){5} \Text(145,110)[l]{$H^-$}
        \DashLine(100,50)(140,50){5} \Text(145,50)[l]{$H^+$}

        \Gluon(210,50)(250,50){4}{4} \Text(205,50)[r]{$g$} 
        \Vertex(250,50){1}
        \Gluon(210,110)(250,110){4}{4} \Text(205,110)[r]{$g$}
        \Vertex(250,110){1}
        \ArrowLine(250,110)(250,50) \Text(245,80)[r]{b}
        \ArrowLine(250,50)(310,50) \Text(280,45)[t]{b}
        \ArrowLine(310,50)(310,110) \Text(315,80)[l]{t}
        \ArrowLine(310,110)(250,110) \Text(280,115)[b]{b}
        \Vertex(310,50){1}
        \Vertex(310,110){1}
        \DashLine(310,110)(350,110){5} \Text(355,110)[l]{$H^+$}
        \DashLine(310,50)(350,50){5} \Text(355,50)[l]{$H^-$}

        \Gluon(210,170)(250,170){4}{4} \Text(205,170)[r]{$g$} 
        \Vertex(250,170){1}
        \Gluon(210,230)(250,230){4}{4} \Text(205,230)[r]{$g$}
        \Vertex(250,230){1}
        \ArrowLine(250,230)(250,170) \Text(245,200)[r]{t}
        \ArrowLine(250,170)(310,170) \Text(280,165)[t]{t}
        \ArrowLine(310,170)(310,230) \Text(315,200)[l]{b}
        \ArrowLine(310,230)(250,230) \Text(280,235)[b]{t}
        \Vertex(310,170){1}
        \Vertex(310,230){1}
        \DashLine(310,230)(350,230){5} \Text(355,230)[l]{$H^-$}
        \DashLine(310,170)(350,170){5} \Text(355,170)[l]{$H^+$}
    \end{picture} \\
    \vspace*{-1cm}
    \caption{\it The $s$-channel triangle diagrams and box diagrams
      contributing to $gg\to H^+H^-$.}
    \end{center} 
    \end{figure}
    
    When the form factors are evaluated numerically, the masses and
    couplings are consistently used in two-loop order \cite{Carena}.

\section{Charged Higgs Pairs in $gg$ collisions}
Two mechanisms contribute to the production of charged Higgs-boson
pairs in $gg$ fusion, exemplified by the generic diagrams in Fig.~3:
(i) Virtual neutral CP-even Higgs bosons $h^0,H^0$, which subsequently
decay into $\hp\hm$ final states, are coupled to gluons by $t,b$ quark
triangles; (ii) The coupling between charged Higgs bosons and gluons
is also mediated by heavy-quark box diagrams. As a result of CP
invariance, the neutral CP-odd Higgs boson $A^0$ does not couple to
pairs of charged Higgs bosons. The $Z$ boson cannot mediate the
coupling either: The vector component of the wave function does not
couple to the initial $gg$ state as a result of the Landau--Yang
theorem, the CP-odd scalar component does not couple to the CP-even
$\hp\hm$ state.  Forbidden by the Landau--Yang theorem, virtual
photons cannot contribute either.

  In the \underline{triangle diagrams} of Fig.~3 the
  gluons are coupled
  to the spin $S_z = 0$ along the collision axis. The transition
  matrix element associated with this mechanism can therefore be expressed
  by the product of one form factor $F^Q_\triangle$, depending on the scaling
  variable $\tau_Q = 4m_Q^2/\hat s$, with the generalized coupling
  $C^Q_\triangle$ defined as
  \begin{equation}
    C^Q_\triangle = \sum_{H_i=h^0,H^0} \lambda_{H^+H^-H_i} \frac{m_Z^2}
                    {\shat-m_{H_i}^2+i m_{H_i} \Gamma_{H_i}} g_Q^{H_i} \; .
  \end{equation}
  The couplings $g_Q^{H_i}$ denote the Higgs--quark couplings in units of the
  SM Yukawa couplings, collected in Table 1; the Higgs
  self-couplings $\lambda_{\hp\hm H_i}$ have
  been
  defined in Eqs.\,(\ref{eq:lambda1}),(\ref{eq:lambda2}). $\sqrt{\hat s}$ is
  the
  $gg$ center-of-mass energy. For the
  sake of convenience, the well-known triangle form
  factor $F^Q_\triangle (\tau_Q)$ is
  recapitulated in Appendix 1. In the limit of large [$t$] and small [$b$]
  quark-loop masses, the form factor $F^Q_\triangle$ simplifies considerably
  ($\tau_t \equiv 4m_t^2/\shat$, $\tau_b \equiv 4m_b^2/\shat$):
  \begin{eqnarray}
      F^t_\triangle (\tau_t \gg 1) & \to & \frac{2}{3} \\
      F^b_\triangle (\tau_b \ll 1) & \to & -\frac{\tau_b}{4}{\left[
      \log \frac{\tau_b}{4} +i\pi \right]}^2 \; .
  \end{eqnarray}
  These expressions provide useful approximations in practice for a large 
  range of Higgs masses.
  
  The \underline{box diagrams} in Fig.~3 contribute to the two spin
  states $S_z=0$ and 2. The tensor basis for the states is given
  explicitly in Appendix 1. For $S_z=2$ two independent elements can
  be defined, carrying positive and negative parity under space
  reflection.  For $S_z=0$ only the basis element carrying positive
  parity is realized since the [$S_z=0;\:P=-$] element is odd under CP
  transformation and thus forbidden for CP conserving
  $gg\rightarrow\hp\hm$ transitions.  Therefore we are left with three
  independent form factors $F,G$ and $H$, corresponding to the
  spin-C/P states $0^{++},2^{++}$ and $2^{--}$. It turns out to be
  convenient to split each form factor into three components, which
  are proportional to ${\text{tan}}^2\beta$ and ${\text{cot}}^2\beta$,
  and independent of $\tanb$:
  \begin{alignat}{3}
           F_\Box &= {\text{tan}}^2\beta F_\Box^1
                    +{\text{cot}}^2\beta F_\Box^2
                    +F_\Box^3 \\
           G_\Box &= {\text{tan}}^2\beta G_\Box^1
                    +{\text{cot}}^2\beta G_\Box^2
                    +G_\Box^3 \\
           H_\Box &= {\text{tan}}^2\beta H_\Box^1
                    +{\text{cot}}^2\beta H_\Box^2
                    +H_\Box^3 \; .
  \end{alignat}  
  The form factors depend on the square of the invariant energy $\shat$ and the
  momentum transfer squared $\that,\uhat$. These Mandelstam
  variables are given by the gluon beam energy $\hat{E}$ and the
  $H^\pm$ production angle $\hat{\theta}$ with respect to the
  $gg$ axis in the partonic c.m.\ system:
  \begin{alignat}{3}
      \shat &= 4 {\hat{E}}^2 \notag \\
      \that &= m_{\hp}^2-2 {\hat{E}}^2
               \left(1-\bhat
               \,\text{cos}\hat{\theta}\right) \\
      \uhat &= m_{\hp}^2-2 {\hat{E}}^2
               \left(1+\bhat
               \,\text{cos}\hat{\theta}\right)
      \; \notag
  \end{alignat}
  with the velocity $\bhat=(1-4 m_{H^\pm}^2/\shat)^{1/2}$. Compact analytical
  expressions in terms of these Mandelstam variables are given in
  Appendix 1.

In the double limit of large $t$ mass and small $b$ mass, the form factors can
be reduced to very simple expressions:
\begin{eqnarray}
F_\Box^1 & \to & 0 \label{eq:flimit} \nonumber \\
F_\Box^2 & \to & \frac{\tau_b}{4}
                 {\left[ \log\frac{\tau_b}{4} +i\pi\right]}^2  
                 +\frac{2}{3} \nonumber \\
F_\Box^3 & \to & \frac{\tau_b}{2}
                 {\left[ \log\frac{\tau_b}{4} +i\pi\right]}^2  \\[2mm]
G_{\Box} & \to & 0 \nonumber \\
H_{\Box} & \to & 0 \nonumber \; .
\end{eqnarray}

As expected, the $S_z=2$ form factors $G_\Box$ and $H_\Box$ vanish in
the limit of large top masses. The non-zero components of the $S_z=0$
form factor $F_\Box$ can be interpreted in a transparent way.
Pinching the top propagator in the $bbbt$ box effectively leads to a
bottom-quark triangle, which is accounted for by the $\tau$-dependent
expressions in $F_\Box$. The remaining constant term $2/3$ corresponds
to the sum of the asymptotic limits of the $bttt$ box [$=-1/3$] and
the contribution generated by the $bbtt$ box [$=+1$].

  The form factors are compared with their asymptotic values in Fig.\,4
  for a particular value of the scattering angle.
  The partonic center-of-mass energy has been chosen near
  threshold, $\sqrt{\shat}=310\,$GeV for $m_{H^\pm}=150$\,GeV, where
  the dominant contributions to the hadron cross section are generated. The
  imaginary part of $F_\triangle$ is very small and is therefore not shown
  in the figure. Apparently, the box form factors approach the asymptotic
  values only for large quark masses. For $m_t=175\,$GeV the asymptotic box
  expressions are not yet useful in practice. Notably the real part
  of $F_\Box$, which for most of the parameter space gives the leading
  contribution to the cross section, approaches asymptotia only very slowly.
  Nevertheless, the comparison in the asymptotic parameter range provides
  an important cross check of the
  general calculation for realistic $b,t$ quark masses.
  \begin{figure}[ht]
  \begin{center}
  \epsfig{file=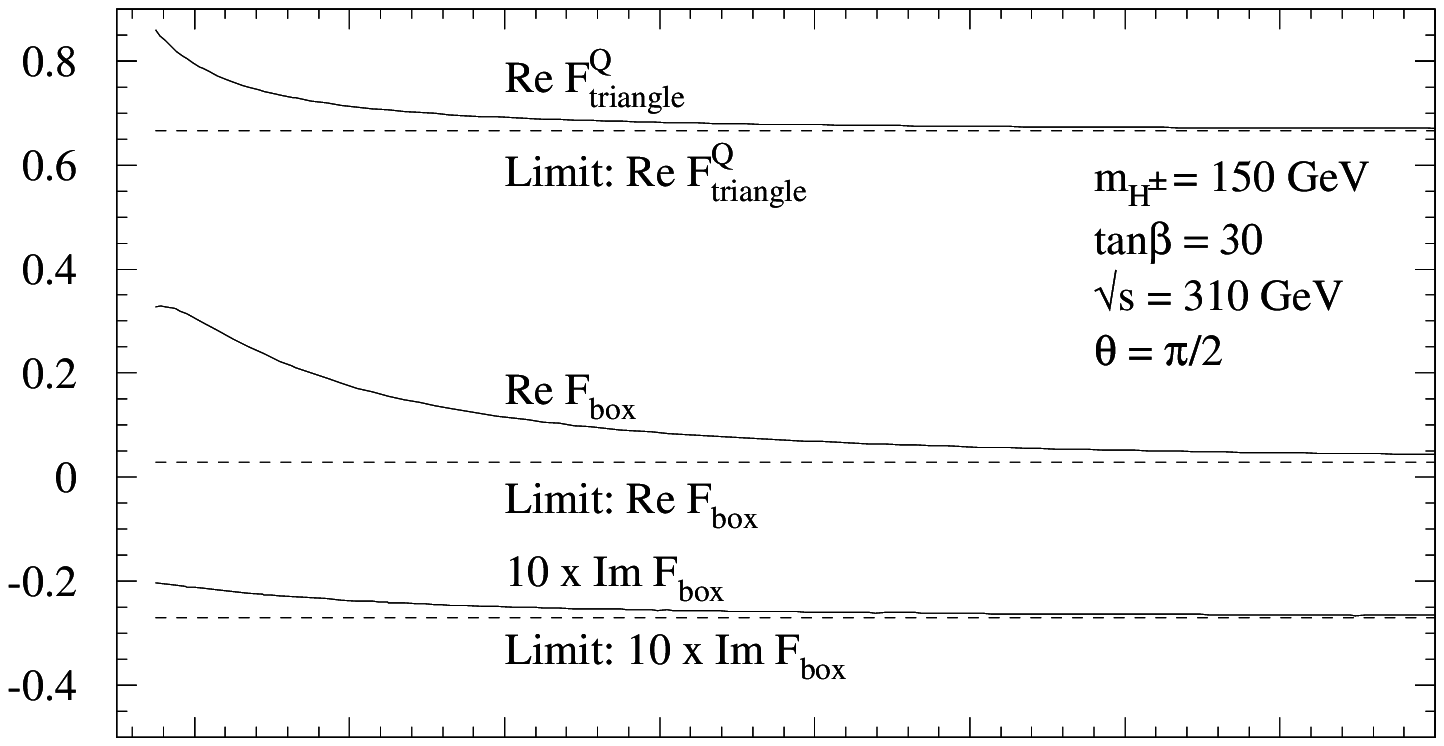,width=14cm}
  \epsfig{file=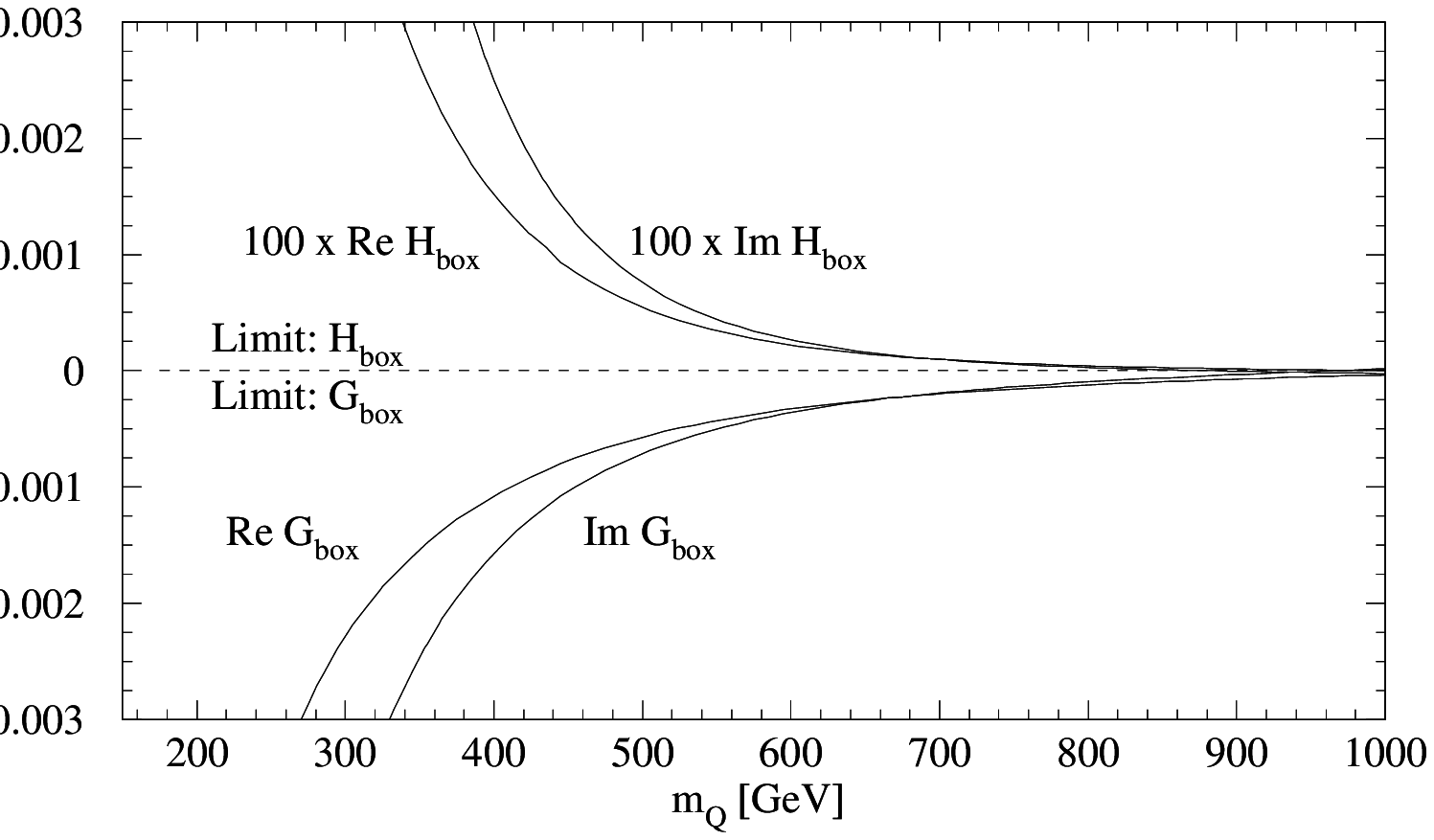,width=14cm}
  \caption{\it The form factors $F_\triangle$, $F_\Box$ and $G_\Box$,
               $H_\Box$ in comparison with their asymptotic limits
               for large quark-loop masses $Q\leftarrow t$.} 
  \end{center}
  \end{figure}
  \clearpage

  \begin{figure}[p]
  \begin{center}
  \epsfig{file=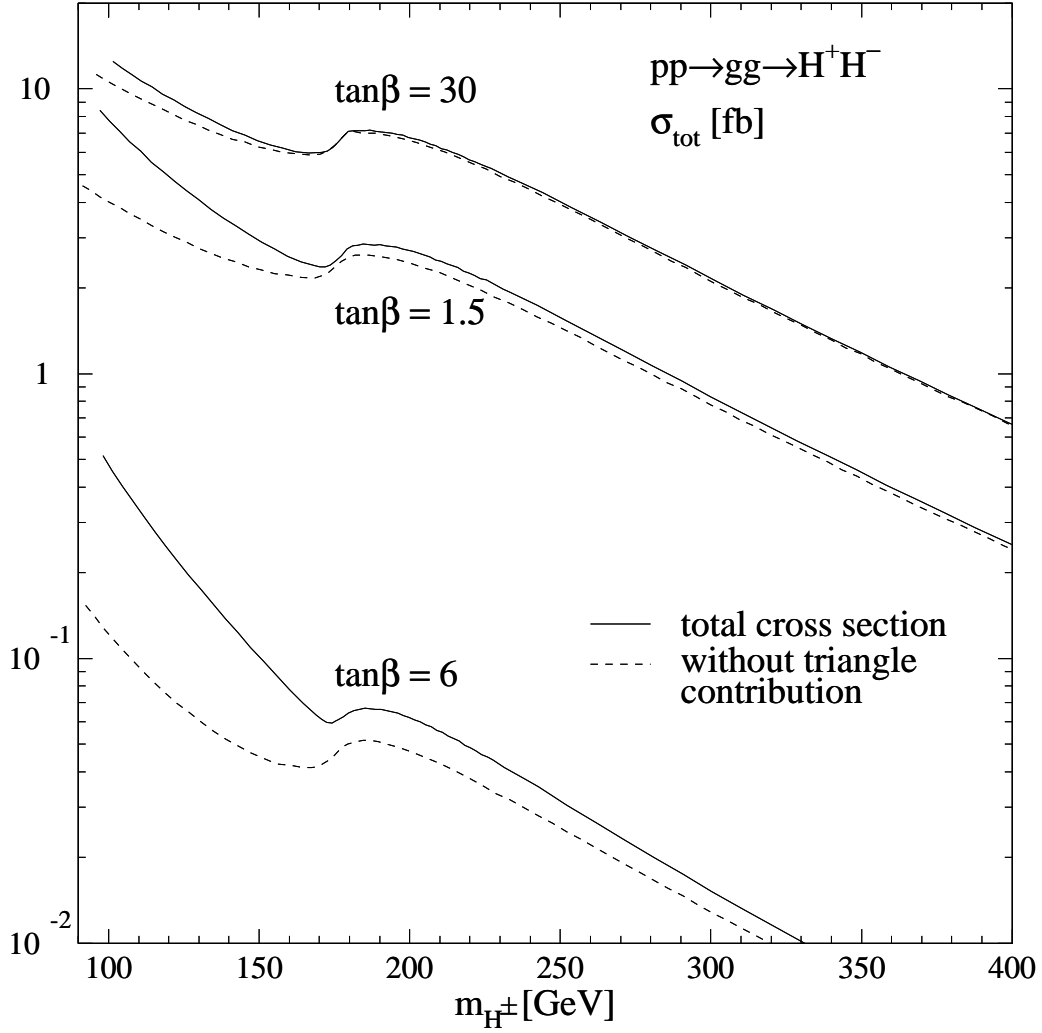,width=14cm,angle=0}
  \caption[]{{\it Total cross sections (full lines)
               and the contributions without the triangle
               diagrams (dashed lines)
               for gluon fusion of charged Higgs-boson
               pairs at the LHC} [{\sl CTEQ4M parton densities \cite{CTEQ}
               with
               $\alpha_s \left( m_Z^2\right) =0.116$}].} 
  \end{center}
  \end{figure}

The differential cross section at the parton level can
        finally be written in
        the form
\begin{equation}
  \frac{d\hat{\sigma}}{d\hat t} [gg\to H^+H^-] = 
  \frac{G_F^2\alpha_s^2}{256 (2\pi)^3}
  \Big[ \big| \sum_{Q=t,b} C^Q_\triangle F^Q_\triangle + F_\Box \big|^2
+ \left| G_\Box \right|^2 + \left| H_\Box \right|^2 \Big]
  \label{eq:cxnpart} \; .
\end{equation}
The hadronic cross section for the process $pp\to gg\to H^+H^-$ can
finally be derived
by integrating (\ref{eq:cxnpart}) over the production angle $0\le \hat\theta
\le \pi$ and the $gg$ luminosity:
\begin{equation}
\sigma[pp\to gg\to H^+H^-] = \int_{4m_{H^\pm}^2/s}^1 d\tau
\frac{d{\cal L}^{gg}}{d\tau} \hat\sigma (\hat s = \tau s) \; .
\end{equation}
It is likely that QCD corrections to gluon-fusion processes, known for
the triangle diagrams, unknown still for the box diagrams, enhance the
event rate (cf. Ref.\cite{Spira}). In parallel to other gluonic
processes, the corrections are expected to remain under control if the
factorization scale in the $gg$ luminosity is chosen to be of the
order of the invariant $(\hp\hm)$ energy.

  \begin{figure}[h]
  \begin{center}
  \epsfig{file=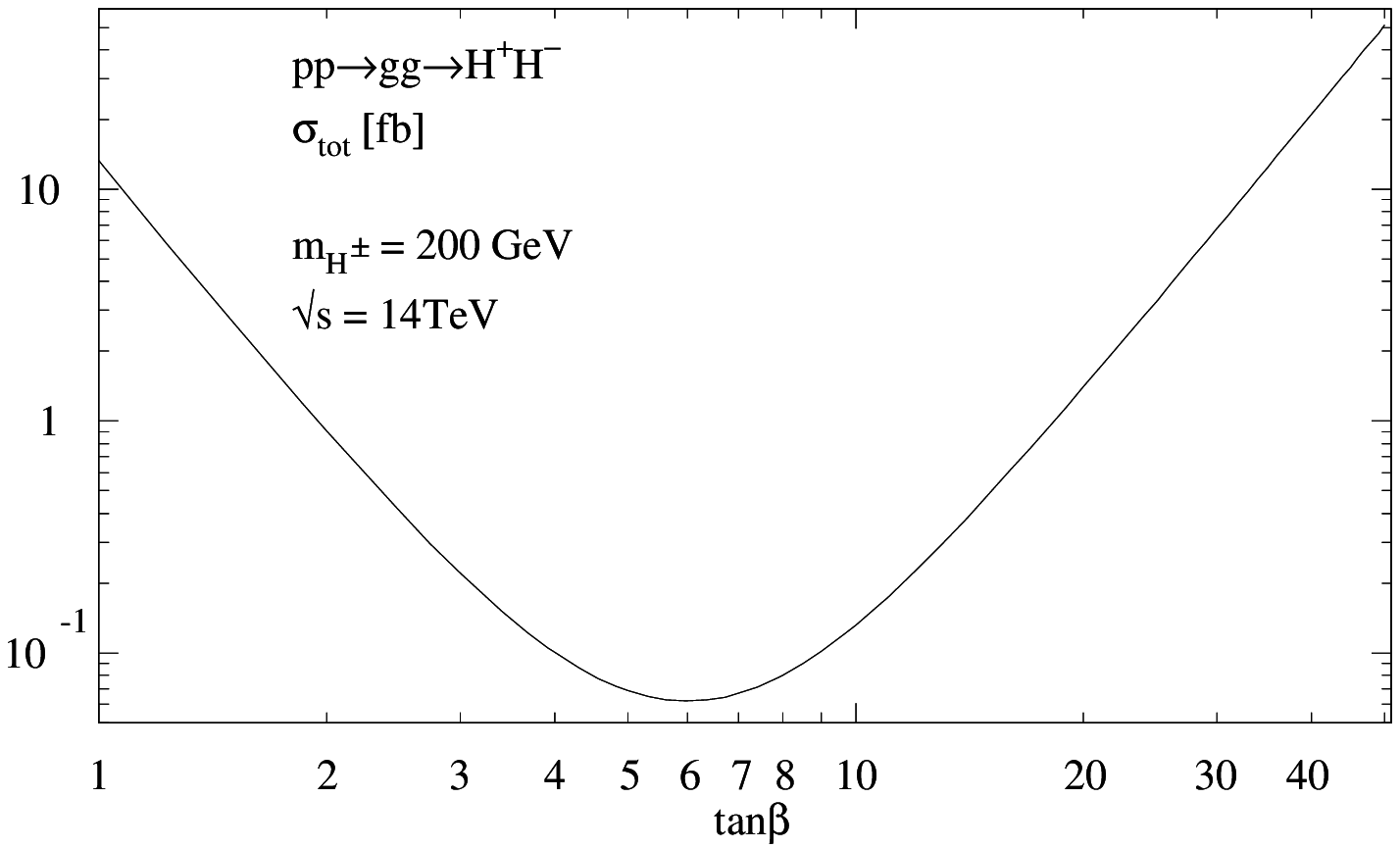,width=14cm,angle=0}
  \caption{\it Dependence of the gluon-fusion cross section of
               charged Higgs-boson pairs on the parameter $\tanb$.} 
  \end{center}
  \end{figure}
  
  We have evaluated the $pp$ cross section
  numerically\footnote{FORTRAN code in Ref.\ \cite{Krause}.}  for the
  LHC energy $\sqrt{s} = 14\,$~TeV.  The (pole) mass of the top quark
  has been set to $m_t=175$ GeV and the bottom quark mass to
  $m_b=5\,$GeV. The results are shown for three values $\tanb=1.5,6$
  and 30 in Fig.~5 as a function of the charged Higgs boson mass. [The
  masses of the neutral Higgs bosons $h^0$ and $H^0$ which enter
  through the charges $C^Q_\triangle$, are fixed by $m_{H^\pm}$ and
  $\tanb$ in the MSSM.]  For small values of $\tanb$ and modest values
  of $m_{H^\pm}$, the triangle diagram provides the dominant
  contribution. With rising Higgs mass and for large $\tanb$, the box
  diagrams become increasingly important; they dominate for Higgs
  masses above the top mass. The choice $\tanb = 6$ in Fig.~5
  corresponds to the minimal value of the cross section. This is
  demonstrated in Fig.~6 for a fixed Higgs mass $m_{H^\pm}=200$ GeV
  and $\tanb$ varied between 1 and 50.  The minimal cross section is
  realized at $\tanb\sim \sqrt{m_t/m_b} \sim 6$, roughly equivalent to
  the minimum of the average $QQH^\pm$ coupling $\langle g_{QQH^\pm}
  \rangle = {\left( m_t^2 \tan^2\!\beta + m_b^2 \cot^2\!\beta
    \right)}^{1/2}$.
  
  The cross section for gluon fusion of charged Higgs boson pairs is
  smaller than the cross section for the Drell--Yan mechanism. In the
  interesting mass range $m_{H^\pm} > 200\,$~GeV, the ratio is about
  $0.4$ and $0.2$ for $\tanb =30$ and $1.5$ respectively. It is
  noteworthy that the angular distributions of the charged Higgs
  bosons are different for the two mechanisms. While the Drell--Yan
  mechanism generates a $\sin^2\!\theta$ distribution in the
  $(\hp\hm)$ center-of-mass frame, the gluon-fusion processes are not
  suppressed by angular-momentum conservation in the forward
  direction; for example, the amplitudes built up by the Higgs
  exchange diagrams, are isotropic in $\theta$. Even though the
  gluon-fusion mechanism is not dominant for the production of charged
  Higgs-boson pairs, it is nevertheless of great interest since the
  $gg$ fusion cross section and the angular distribution are affected
  by the size of the trilinear self-couplings between the CP-even
  neutral and charged Higgs bosons.

\section{Conclusions}
As shown in the representative examples of Fig.~5, the cross sections
for charged Higgs masses below 300 GeV vary between 1 and 10\,fb in a
large part of the MSSM parameter space. This corresponds to rates of
100 and 1,000 events for an integrated luminosity of $\int {\cal L} =
100\,\text{fb}^{-1}$, which is expected in a year of high-luminosity
runs at the LHC. For small Higgs masses the triangle contribution
appears large enough to generate a significant transition amplitude;
this will enable us to measure the trilinear self-couplings
$\lambda_{H^+H^-h^0}$ and $\lambda_{H^+H^-H^0}$ of the CP-even neutral
and charged Higgs bosons. Charged Higgs particles decay predominantly
to $\tau\nu_\tau$ final states at large $\tanb$, leading to asymmetric
$\tau^+\tau^-$ configurations in the decays of the two Higgs
particles. For small $\tanb$, the potentially dominant
chargino/neutralino decays are more difficult to control
experimentally. These experimental problems must be approached,
however, in dedicated detector and background simulations, which are
beyond the scope of this first-step theoretical analysis.  \newline
\newline
{\bf Acknowledgement:}\\
We thank V.A.\ Smirnov for valuable discussions on the heavy-mass
expansion of Feynman diagrams.

\newpage
\section*{APPENDICES}
\subsection*{1.\,\sl Form factors in $gg \rightarrow \hp\hm$}

  \underline{\sl Parameter definitions}: 
  \begin{alignat*}{3}
      \hat s & = (p_a+p_b)^2, \qquad \hat t & = (p_a - p_c)^2,
      \qquad \hat u & = (p_b - p_c)^2 \qquad\qquad\qquad\qquad \\
      \hat t_1 & = \hat t - m_{H^\pm}^2, \qquad \hat u_1 & = \hat u -
      m_{H^\pm}^2, \qquad p_t^2 & = \hat{t_1}\hat{u_1}/{\hat s} -
      m_{H^\pm}^2 \\
      \tau_Q & = 4 m_Q^2/\shat, \qquad \quad
      \hat{\beta} & = \left(1-4 m_{H^\pm}^2/\shat\right)^{1/2}
  \end{alignat*}
  \underline{\sl Scalar integrals} \cite{Veltman}:
  \begin{alignat*}{3}
      C^{\alpha\beta\gamma}_{ij} &=
      \int \frac{d^4 q}{i \pi^2}
      \frac{1}{\left[ q^2 - m_\alpha^2 \right]
               \left[ {\left( q+p_i \right)}^2 - m_\beta^2 \right]
               \left[ {\left( q+p_i+p_j \right)}^2 - m_\gamma^2 \right]}
      \\
      D^{\alpha\beta\gamma\delta}_{ijk} &=
      \int \frac{d^4 q}{i \pi^2}
      \frac{1}{\left[ q^2 - m_\alpha^2 \right]
               \left[ {\left( q+p_i \right)}^2 - m_\beta^2 \right]
               \left[ {\left( q+p_i+p_j \right)}^2 - m_\gamma^2 \right]
               \left[ {\left( q+p_i+p_j+p_k \right)}^2 - m_\delta^2 \right]}
      \\ \\ 
      &\phantom{sss}\text{with}\phantom{sss}
      \alpha,\beta,\gamma,\delta = t,b \,  \quad \text{and} \quad
      i,j,k = 1,\ldots,4
      \\ 
      &\phantom{ssswithsss}
                     p_1 = p_a \, , \; p_2 = p_b \, , \;
                     p_3 = - p_c \, , \; p_4 = - p_d
                     \;\;\;\text{in}\;\;\;
                     a+b \rightarrow c+d 
  \end{alignat*}  
  \vspace{3mm}
  \underline{\sl Tensor basis}:
  \begin{alignat*}{3}
    P=+\qquad &S_z = 0 \, : \quad
      A_1^{\mu\nu} = g^{\mu\nu} 
                    - \frac{p_a^\nu p_b^\mu}
                           {\left( p_a p_b \right)} \\
      &S_z = 2 \, : \quad
      A_2^{\mu\nu} = g^{\mu\nu} 
                    + \frac{p_c^2 p_a^\nu p_b^\mu}
                           {p_t^2 \left( p_a p_b \right)}
                    - \frac{2 \left( p_b p_c \right) p_a^\nu p_c^\mu}
                           {p_t^2 \left( p_a p_b \right)}
                    - \frac{2 \left( p_a p_c \right) p_b^\mu p_c^\nu}
                           {p_t^2 \left( p_a p_b \right)}
                    + \frac{2 p_c^\mu p_c^\nu}
                           {p_t^2} \\
    P=-\qquad &S_z = 0 \, : \quad
    \bigg[ A_3^{\mu\nu} = \frac{1}{\left( p_a p_b \right)}
                      \epsilon^{\mu\nu p_a p_b} 
      \;\text{: CP odd }\Rightarrow \text{ not coupling to } \hp\hm\,\bigg]
          \\
      &S_z = 2 \, : \quad
      A_4^{\mu\nu} = \frac{p_c^\mu \epsilon^{\nu p_a p_b p_c}
                         +  p_c^\nu \epsilon^{\mu p_a p_b p_c}
                         +  \left( p_b p_c \right) \epsilon^{\mu\nu p_a p_c}
                         +  \left( p_a p_c \right) \epsilon^{\mu\nu p_b p_c}}
                           {\left( p_a p_b \right) p_t^2}
  \end{alignat*} 
  \qquad normalization and orthogonality:
  $A_i A_j = 2 \delta_{ij}$ \\ \\
  \underline{\sl Matrix element}:
  \begin{alignat*}{3}
      {\cal M} \left( g_a g_b \rightarrow H^+ H^- \right) &=
      {\cal M}_\triangle + {\cal M}_\Box \\
      {\cal M}_\triangle &= \frac{G_F \alpha_s \hat{s}}
                                       {2 \sqrt{2} \pi}
                            \sum_{Q=t,b} C_\triangle^Q F^Q_\triangle
                                  A_{1\mu\nu}
                                  \epsilon_a^\mu \epsilon_b^\nu
                                  \delta_{ab} \\
      {\cal M}_\Box &= \frac{G_F \alpha_s \hat{s}}
                            {2 \sqrt{2} \pi}
                       \left(  F_\Box A_{1\mu\nu}
                              +G_\Box A_{2\mu\nu}
                              +H_\Box A_{4\mu\nu} \right)
                                  \epsilon_a^\mu \epsilon_b^\nu
                                  \delta_{ab}
      \qquad\qquad
  \end{alignat*}
  \underline{\sl Triangle form factor}:
  \begin{alignat*}{3}
      F^Q_\triangle &= \tau_Q \left[ 1+\left( 1-\tau_Q \right) 
                                    f \left( \tau_Q \right)
                             \right] \\
      &\phantom{=s} f \left( \tau_Q \right) = 
     \left\{  
       \begin{array}{rl} 
                  -\frac{1}{4}
                  {\left[ \text{log} \left(
                         \frac{1+\sqrt{1-\tau_Q}}
                              {1-\sqrt{1-\tau_Q}} \right)
                                 -i \pi \right]}^2 \;\, 
                  & :\;\; \tau_Q \le 1 \\
                  &              \\ 
                  {\text{arcsin}}^2 \frac{1}{\sqrt{\tau_Q}}
                  \qquad\qquad\qquad\quad 
                  & :\;\; \tau_Q > 1
       \end{array} 
      \right.  \qquad\qquad\qquad\quad
  \end{alignat*}
  \underline{\sl Box form factors}:
  \begin{small}
  \begin{alignat*}{3}
    F_\Box^1 = \frac{2 m_b^2}{\shat} &\bigg[
                 \left(1+2m_t^2C_{12}^{ttt}\right)
              + \left( m_t^2+m_b^2-m_{H^\pm}^2 \right)
              \Big\{ -\frac{\thone}{\shat} 
                         C_{13}^{ttb}
                       -\frac{\uhone}{\shat}
                         C_{23}^{ttb}
              +m_t^2 \left( D_{123}^{tttb}+D_{213}^{tttb} \right)
              \qquad\qquad\qquad\qquad\qquad\qquad \\
              &+\frac{1}{2}\left( m_t^2+m_b^2+p_t^2 \right)
                                    D_{132}^{ttbb}
                 \Big\} 
                 \bigg]
               +\, \frac{2 m_b^2}{\shat} \bigg[b \leftrightarrow t \bigg]   
  \end{alignat*}
  \begin{equation*}
    F_\Box^2 = F^1_\Box \left( m_t \leftrightarrow m_b \right)
    \qquad\qquad\qquad\qquad\qquad\qquad\qquad\qquad\qquad\qquad\quad 
    \qquad\qquad\qquad\qquad\qquad\qquad
  \end{equation*}
  \begin{alignat*}{3} 
    F_\Box^3 = 2m_b^2m_t^2 
     &\bigg[  -\frac{4\hat{t}_1}{\hat{s}^2}
                C_{13}^{ttb}
              -\frac{4\hat{u}_1}{\hat{s}^2}
                C_{23}^{ttb}
           +\Big( \frac{4m_t^2}{\hat{s}}-1 \Big)
            \left( D_{123}^{tttb}+D_{213}^{tttb} \right)
           +\Big\{ \frac{2}{\hat{s}} \left( m_t^2+m_b^2+p_t^2 \right)
                   -1 \Big\}
                   D_{132}^{ttbb} \bigg] \\
          +\, 2m_b^2m_t^2 &\bigg[ b \leftrightarrow t \bigg]
       \qquad\qquad\phantom{s}
  \end{alignat*}
  \begin{alignat*}{3}
    G_\Box^1 = \frac{m_b^2}{p_t^2} \bigg[
               &\Big\{ \frac{\hat{s}}{2}-p_t^2
                        +M^4_{bt} \Big\} C_{12}^{bbb}
              + N( \thone )C_{13}^{ttb} 
              +  N( \uhone )C_{23}^{ttb}
              +\Big( \frac{\hat{s}}{2}+m_b^2+m_t^2-m_{H^\pm}^2 \Big)
                        \Big( {\hat{\beta}}^2
                       -\frac{2p_t^2}{\hat{s}} \Big) 
                         C_{34}^{tbt} \; \\
              +&\Big\{M^6+L_{bt} (\thone,\uhone )\Big\}
                 D_{123}^{bbbt}
              + \Big\{M^6+L_{bt} (\uhone,\thone )\Big\}
                 D_{213}^{bbbt}
              + \Big\{ M^6 +\left( m_b^4+m_t^4 \right)
                 \Big( \frac{1}{2}
                       +\frac{p_t^2-m_{H^\pm}^2}{\hat{s}} \Big) \\
               +&\left( m_b^2+m_t^2 \right) p_t^2
                        \Big(  \frac{1}{2}
                              -\frac{m_{H^\pm}^2}{\hat{s}} \Big) \Big\}
               D_{132}^{ttbb}
               \bigg]
              +\,\frac{m_b^2}{p_t^2}\bigg[
                  b \leftrightarrow t \bigg]
  \end{alignat*}
  \begin{alignat*}{3}
      \phantom{G_\Box^1 =s}\text{where:}\quad
      &M^4_{bt} = \frac{2}{\shat}\Big[
                       {\big( m_t^2-m_{H^\pm}^2 \big)}^2-m_b^4 \Big]
                       + 2\left( m_t^2-m_{H^\pm}^2 \right) \\
      &M^6 = \frac{m_t^6-m_t^4 m_b^2-m_t^2 m_b^4+m_b^6}{\shat}
             +m_b^2m_t^2\Big( 2\frac{\thone\uhone}{{\shat}^2}
                                                    -1 \Big) \\
      &N(\thone) = \frac{\thone}{\shat}
                   \Big[ 2\big( m_{H^\pm}^2-m_b^2-m_t^2 \big)
                   \frac{\that}{\shat}
                  +\frac{{\thone}^2}{\shat} \Big] \\
      &L_{bt}(\thone,\uhone) = 
        m_b^4\left[ \frac{1}{2}+\frac{\uhone}{\shat}
                             \Big( 1-\frac{\uhone}{\shat} \Big)
                             +\frac{p_t^2}{\shat} \right]
       +m_t^4\left[ \frac{1}{2}
          -\frac{2\uhat+m_{H^\pm}^2}{\shat} \right] \\
      &\qquad\qquad\;\, +m_b^2\left[ \frac{{\bhat}^2 p_t^2}{2}
          -\frac{m_{H^\pm}^4}{\shat} \right]
       +m_t^2\left[ \frac{\uhat}{\shat}
                    \left( 2\uhat+\that \right)
                   -\frac{p_t^2}{2} \right]
                   -\frac{\uhat}{\shat}
                    \left[ \frac{{\uhone}^2}{2}
                   +\uhat\,m_{H^\pm}^2 \right]
      \qquad\qquad\qquad
  \end{alignat*}

  \begin{equation*}
    G_\Box^2 = G_\Box^1 \left( m_t \leftrightarrow m_b \right)
    \qquad\qquad\qquad\qquad\qquad\qquad\qquad\qquad\qquad
    \qquad\qquad\qquad\qquad\qquad\qquad\qquad\qquad\, 
  \end{equation*}
  \begin{alignat*}{3}  
    G_\Box^3 = \frac{4 m_t^2 m_b^2}{\hat{s} p_t^2} &\bigg[
                  - \left( 2m_t^2-2m_b^2+2m_{H^\pm}^2-\hat{s} \right)
                    C_{12}^{ttt}
                  - \frac{2 \hat{t}_1 \hat{t}}{\hat{s}}
                    \, C_{13}^{ttb} 
                  - \frac{2 \hat{u}_1 \hat{u}}{\hat{s}}
                    \, C_{23}^{ttb}
                  + \left( \hat{s}-4m_{H^\pm}^2-2p_t^2 \right)
                    \, C_{34}^{tbt} \\
                 &\;+ L_{bt}(\uhat)
                    D_{123}^{tttb} 
                  + L_{bt}(\that)
                    D_{213}^{tttb} 
                  + \Big\{ {\left( m_t^2-m_b^2 \right)}^2
                        + \left( m_t^2+m_b^2 \right) p_t^2 \Big\}
                                 D_{132}^{ttbb}
                  \bigg] 
               +\,\frac{4 m_t^2 m_b^2}{\hat{s} p_t^2}
                \bigg[ b \leftrightarrow t \bigg]
         \qquad\qquad\qquad\qquad\qquad\qquad\qquad\quad
  \end{alignat*}
  \begin{alignat*}{3}
       \phantom{G_\Box^1 =s}\text{where:}\quad
     L_{bt}(\uhat) &=  {\left( m_t^2-m_b^2 \right)}^2
                        + \hat{u}\left( \hat{u}-2m_b^2 \right)
                        - 2m_t^2 \hat{u}_1^2/\hat{s}
                       \qquad \text{and} \quad \uhat \Rightarrow \that
     \qquad\qquad\qquad\qquad\qquad\quad
  \end{alignat*}
  \begin{alignat*}{3}  
     H_\Box^1 = \frac{m_b^2}{p_t^2} &\bigg[
                \left( \hat{t}_1-\hat{u}_1 \right)
                        M_{bt}^2 \, 
                  C_{12}^{ttt} 
                + \frac{\hat{t}_1^2}{\hat{s}^2}
                          \left( t+m_{H^\pm}^2 \right) 
                  \, C_{13}^{ttb}
                 - \frac{\hat{u}_1^2}{\hat{s}^2}
                          \left( u+m_{H^\pm}^2 \right) 
                  \, C_{23}^{ttb}
                + \Big( \frac{\hat{u}_1-\hat{t}_1}{2} \Big)
                          {\hat{\beta}}^2 
                  \, C_{34}^{btb} \\
                &\,+ R_{bt}(\thone,\uhone) D_{123}^{bbbt}     
                - R_{bt}(\uhone,\thone) D_{213}^{bbbt}  
                + \Big( \frac{\hat{u}_1-\hat{t}_1}{2\hat{s}} \Big)
                          \Big\{ {\Big( m_t^2-m_b^2 \Big)}^2
                                +\left( m_t^2+m_b^2 \right) p_t^2 \Big\}
                                 D_{132}^{ttbb} 
                  \bigg] \\
                + \frac{m_b^2}{p_t^2} &\bigg[ b \leftrightarrow t \bigg]
  \end{alignat*} 
  \begin{alignat*}{3}
       \phantom{G_\Box^1 =s}\text{where:}\quad
      &R_{bt}(\thone,\uhone) = \frac{1}{2}
                       \left( \hat{u}+m_{H^\pm}^2 \right)
                       \frac{\hat{u}\hat{u}_1}{\hat{s}}
                      +\frac{1}{2}\left(
                                  \frac{\hat{u}_1-\hat{t}_1}{\hat{s}}
                                  \right)
                       {\left( m_t^2-m_b^2 \right)}^2 \\
                     &\phantom{ssssssssss}
                      +m_b^2\left[ \frac{\hat{u}_1^2}{\hat{s}^2}
                                   \left( \hat{t}_1-\hat{u}_1 \right)
                                  +\frac{p_t^2}{2} \right]
                      +m_t^2\left[ \frac{\hat{u}}{\hat{s}}
                                   \left( \hat{t}_1-\hat{u}_1 \right)
                                  -\frac{p_t^2}{2} \right] \qquad \\
      &M^2_{bt} = 
       \frac{m_{H^\pm}^2+m_t^2-m_b^2}{\hat{s}}-\frac{1}{2}
      \qquad\qquad\qquad\qquad\qquad\qquad\qquad\qquad\qquad\qquad
      \qquad\qquad\qquad
  \end{alignat*}

  \begin{equation*}
     H_\Box^2 = -\, \frac{m_t^2}{m_b^2} \, H_\Box^1 
    \qquad\qquad\qquad\qquad\qquad\qquad\qquad\qquad\qquad
    \qquad\qquad\qquad\qquad\qquad\qquad\qquad\qquad\qquad
  \end{equation*}
  \begin{equation*}
     H_\Box^3 = 0 
    \qquad\qquad\qquad\qquad\qquad\qquad\qquad\qquad\qquad
    \qquad\qquad\qquad\qquad\qquad\qquad\qquad\qquad\qquad\qquad\qquad
  \end{equation*}
  \end{small}

\subsection*{2.\,\sl Pinching top propagators}  
Because of delicate cancellations, the limit $m_t \rightarrow \infty$
is not easy to calculate for the box diagrams involving one $t$ line
and three $b$ lines, etc. We have performed this limit for the
amplitudes first in the Feynman parametrization. The problem however
can also be solved in a systematic way by adopting the scheme for the
expansion in the heavy-quark mass, which has been discussed in
Ref.~\cite{Smirnov}. This method is technically easier than the
Feynman parameter method. The limit is performed by expanding the
scalar integrals inside the form factors; the $m_t$ contributions
arising from the Yukawa couplings are factored out.

The procedure \cite{Smirnov} will be illustrated for a scalar
three-point function including one top propagator:
  \begin{equation*}
     C_{k_1 k_2}^{bbt} = \int\!\frac{d^n\!q}{i\pi^2}
                         \frac{1}{\big[ q^2-m_b^2 \big]     
                         \big[{\left(q+k_1 \right)}^2-m_b^2 \big]
                         \big[{\left(q+k_1+k_2 \right)}^2-m_t^2 \big]}
              \; ,
  \end{equation*}
  which must be expanded up to ${\cal O}((k_ik_j)^2/m_t^6)$. The
  integral is apparently ultraviolet and infrared finite;
  nevertheless, we will work in $n$ dimensions in order to properly
  define the divergences that occur at intermediate steps of the
  calculation. Since the Taylor expansion with respect to the heavy
  quark mass
  \begin{equation*}
     \frac{1}{{\left(q+k_1+k_2 \right)}^2-m_t^2} =
     \underbrace{
     -\frac{1}{m_t^2}-\frac{{\left(q+k_1+k_2 \right)}^2}{m_t^4}}_
                {N_{\,UV}}
     + ...
  \end{equation*}
  leads to ultraviolet divergences of the integral at
  $q\rightarrow\infty$, the expansion cannot be applied in this
  na$\ddot{{\text{\i}}}$ve form.  However, after regularizing the
  divergences in $n$ dimensions, $N_{\,UV}$ can be used in the
  identity
  \begin{equation*}
     \frac{1}{{\left(q+k_1+k_2 \right)}^2-m_t^2} =
     N_{\,UV} + 
     \underbrace{\Big[\frac{1}{{\left(q+k_1+k_2 \right)}^2-m_t^2}
                        -N_{\,UV} \Big]}_{R_{IR}} \; .
  \end{equation*}
  to define a kernel
  \begin{equation*}
  R_{IR} = \frac{(q+k_1+k_2)^2}{m_t^4 \left[(q+k_1+k_2)^2-m_t^2\right]}
  \end{equation*}
  which, together with the two bottom-quark propagators in $C$, can be
  expanded in the variables $(qk_i)/(q^2-m_t^2)$, $(k_ik_j)/m_t^2$ and
  $m_b^2/m_t^2$. After performing the integral of the expansion in $n$
  dimensions, the leading terms generate singularities which exactly
  cancel the singularities of the integral over $N_{\,UV}$. The finite
  pieces and the subsequent terms of the expansion lead to a
  well-behaved series in inverse powers of the heavy quark
  mass. The expansion for this particular example can finally be
  written:
  \begin{alignat*}{3}
     C_{k_1 k_2}^{bbt} &= \frac{1}{m_t^2}
     \left[ 1+\log\left(\frac{m_b^2}{m_t^2}\right)-g(\kappa_b) \right] \\
     &+\frac{1}{m_t^4}\left[ m_b^2 \left( 1+
     2 \log\left(\frac{m_b^2}{m_t^2}\right)-g(\kappa_b)\right)
     + \left(k_1 k_2 +k_2^2 \right)
     \left( \frac{5}{2}+\log\left(\frac{m_b^2}{m_t^2}\right)
           -g(\kappa_b)\right) 
     + \frac{k_1^2}{2} \right] \\
     &+ {\cal O}\left(\frac{(k_ik_j)^2}{m_t^6}\right) \; ,      
  \end{alignat*}
  where $\kappa_b = 4 m_b^2/k_1^2$ and
  \begin{equation*}
     g(\kappa_b) =
     \left\{  
       \begin{array}{rl} 
              2-\sqrt{1-\kappa_b}
              \left[ \log \left\{ \frac{1+\sqrt{1-\kappa_b}}
                                       {1-\sqrt{1-\kappa_b}}
                          \right\}
                      -i \pi \right] \;\, 
                  & :\;\; \kappa_b \le 1 \\
                  &       \\
                  2-2\sqrt{\kappa_b - 1} \arcsin \frac{1}{\sqrt{\kappa_b}}
                  \quad \quad \quad \quad \;\;
                  & :\;\; \kappa_b > 1
       \end{array} 
      \right. 
  \end{equation*}
  
  The same procedure can be applied to the large quark mass limit of
  scalar three-point functions including two top propagators. The
  product of the two top propagators is first
  na$\ddot{{\text{\i}}}$vely Taylor expanded. The top part is then
  split into UV- and IR-divergent contributions as shown before. After
  the Taylor-expansion of the IR-divergent part, the calculation of
  the integrals leads to:
  \begin{alignat*}{3}
     C_{k_1 k_2}^{ttb} = &-\frac{1}{m_t^2}
        -\frac{1}{m_t^4}
         \bigg[ m_b^2\bigg(1+\log\big(\frac{m_b^2}{m_t^2}\big)\bigg)
              \frac{k_1^2}{3}+\frac{k_2^2}{2}+\frac{k_1k_2}{2} \bigg] 
         \qquad\qquad\qquad\qquad \\  
       &+{\cal O}\Big( \frac{(k_ik_j)^2}{m_t^6} \Big) \; .
  \end{alignat*}

  For the scalar three-point function involving three top propagators,
  the na$\ddot{{\text{\i}}}$ve expansion is UV-finite, so that it need
  not be split.  One obtains the following expansion up to the
  required order ${\cal O}((k_ik_j)^3/m_t^8)$:
  \begin{alignat*}{3}
   C_{k_1 k_2}^{ttt} = &-\frac{1}{2}\frac{1}{m_t^2}
        -\frac{1}{m_t^4}\Big[ \frac{k_1k_1+k_1k_2+k_2k_2}{12} \Big] 
         \\
       &-\frac{1}{m_t^6}\Big[ 
         \frac{(k_1^2+k_2^2)k_1k_2}{30}
        +\frac{k_1^2k_2^2}{36}+\frac{(k_1k_2)^2}{45}
        +\frac{k_1^4+k_2^4}{60} \Big] 
        \qquad\qquad\qquad\quad \\
       &+{\cal O}\Big( \frac{(k_ik_j)^3}{m_t^8} \Big) \; .
  \end{alignat*}
  
  The expansion of the four-point functions can be carried out in an
  analogous way. The result is still too long to be presented in plain
  text; yet the analytic expressions are available from \cite{Krause}.

  In contrast to the expansion for large $t$ mass, the subsequent
  expansion for small $b$ mass $m_b^2/\shat \ll 1$ is straightforward. 
  


 \newcommand{\zpc}[3]{{\sl Z. Phys.} {\bf C#1} (19#2) #3}
 \newcommand{\zp}[3]{{\sl Z. Phys.} {\bf #1} (19#2) #3}
 \newcommand{\npb}[3]{{\sl Nucl. Phys.} {\bf B#1} (19#2)~#3}
 \newcommand{\plb}[3]{{\sl Phys. Lett.} {\bf B#1} (19#2) #3}
 \newcommand{\prd}[3]{{\sl Phys. Rev.} {\bf D#1} (19#2) #3}
 \newcommand{\prl}[3]{{\sl Phys. Rev. Lett.} {\bf #1} (19#2) #3}
 \newcommand{\prep}[3]{{\sl Phys. Rep.} {\bf #1} (19#2) #3}
 \newcommand{\fp}[3]{{\sl Fortschr. Phys.} {\bf #1} (19#2) #3}
 \newcommand{\nc}[3]{{\sl Nuovo Cimento} {\bf #1} (19#2) #3}
 \newcommand{\ijmp}[3]{{\sl Int. J. Mod. Phys.} {\bf #1} (19#2) #3}
 \newcommand{\rmp}[3]{{\sl Rev. Mod. Phys.} {\bf #1} (19#2) #3}
 \newcommand{\ptp}[3]{{\sl Prog. Theor. Phys.} {\bf #1} (19#2) #3}
 \newcommand{\sjnp}[3]{{\sl Sov. J. Nucl. Phys.} {\bf #1} (19#2) #3}
 \newcommand{\cpc}[3]{{\sl Comp. Phys. Commun.} {\bf #1} (19#2) #3}
 \newcommand{\mpla}[3]{{\sl Mod. Phys. Lett.} {\bf A#1} (19#2) #3}
 \newcommand{\cmp}[3]{{\sl Commun. Math. Phys.} {\bf #1} (19#2) #3}
 \newcommand{\jmp}[3]{{\sl J. Math. Phys.} {\bf #1} (19#2) #3}
 \newcommand{\nim}[3]{{\sl Nucl. Instr. Meth.} {\bf #1} (19#2) #3}
 \newcommand{\el}[3]{{\sl Europhysics Letters} {\bf #1} (19#2) #3}
 \newcommand{\ap}[3]{{\sl Ann. of Phys.} {\bf #1} (19#2) #3}
 \newcommand{\jetp}[3]{{\sl JETP} {\bf #1} (19#2) #3}
 \newcommand{\jetpl}[3]{{\sl JETP Lett.} {\bf #1} (19#2) #3}
 \newcommand{\acpp}[3]{{\sl Acta Physica Polonica} {\bf #1} (19#2) #3}
 \newcommand{\vj}[4]{{\sl #1~}{\bf #2} (19#3) #4}
 \newcommand{\ej}[3]{{\bf #1} (19#2) #3}
 \newcommand{\vjs}[2]{{\sl #1~}{\bf #2}}
 \newcommand{\hep}[1]{{\sl hep--ph/}{#1}}
 \newcommand{\desy}[1]{{\sl DESY-Report~}{#1}}

\end{document}